\documentclass[11pt]{czipreprint}
\usepackage[utf8]{inputenc}
\usepackage[T1]{fontenc}
\usepackage{graphicx}
\usepackage{amsmath,amssymb}
\usepackage{hyperref}
\usepackage{booktabs}
\usepackage{microtype}
\usepackage{float}
\usepackage{geometry}
\usepackage{siunitx}
\usepackage{xcolor}
\usepackage{subcaption}
\usepackage{longtable}
\usepackage{lineno}
\usepackage[numbers,square,sort&compress]{natbib}

\graphicspath{{img/}{figures/}}

\title{\textbf{A path towards AI-scale, interoperable biological data}}

\author[1]{Brian Aevermann}
\author[2,3,4,5,6,7]{Andrea Califano}
\author[1]{Chi-Li Chiu}
\author[1]{Nathan Clack}
\author[1,8]{William M. Clemons Jr.}
\author[1,26]{Jonah Cool}
\author[1]{Florence D. D'Orazi}
\author[1]{Elizabeth Fahsbender}
\author[9]{Joseph L. DeRisi}
\author[9]{Joshua E. Elias}
\author[10,11,12,13]{Scott E. Fraser}
\author[9]{Carlos G. Gonzalez}
\author[10]{Matthias Haury}
\author[1]{Theofanis Karaletsos}
\author[14,15,16,17]{Shana O. Kelley}
\author[14,18,19,20]{Aly A. Khan}
\author[1]{Alan R. Lowe}
\author[9,21,22,23]{Emma Lundberg}
\author[14]{Ryan A. McClure}
\author[1]{Stephani Otte}
\author[2]{Evan O. Paull}
\author[9]{Lo\"ic A. Royer}
\author[1]{Dana Sadgat}
\author[9,24]{Sandra L. Schmid}
\author[1]{Samantha Scovanner}
\author[1]{Cathy Stolitzka}
\author[1,25]{Jason R. Swedlow}
\author[9]{Joan Wong}
\author[1]{Garabet Yeretssian}
\author[1]{Patricia Brennan}
\author*[1]{Ambrose J. Carr}

\email{acarr@chanzuckerberg.com}

\affiliation[1]{Chan Zuckerberg Initiative, Redwood City, CA, USA}
\affiliation[2]{Chan Zuckerberg Biohub New York, New York, NY, USA}
\affiliation[3]{Department of Systems Biology, Columbia University Irving Medical Center, New York, NY, USA}
\affiliation[4]{Department of Biochemistry \& Molecular Biophysics, Columbia University Irving Medical Center, New York, NY, USA}
\affiliation[5]{Department of Biomedical Informatics, Columbia University Irving Medical Center, New York, NY, USA}
\affiliation[6]{Herbert Irving Comprehensive Cancer Center, Columbia University Irving Medical Center, New York, NY, USA}
\affiliation[7]{Department of Medicine, Vagelos College of Physicians and Surgeons, Columbia University, New York, NY, USA}
\affiliation[8]{Division of Chemistry and Chemical Engineering, California Institute of Technology, Pasadena, CA, USA}
\affiliation[9]{Chan Zuckerberg Biohub San Francisco, San Francisco, CA, USA}
\affiliation[10]{Chan Zuckerberg Imaging Institute, Redwood City, CA, USA}
\affiliation[11]{Department of Biomedical Engineering, University of Southern California, Los Angeles, CA, USA}
\affiliation[12]{Translational Imaging Center, University of Southern California, Los Angeles, CA, USA}
\affiliation[13]{Molecular and Computational Biology Department, University of Southern California, Los Angeles, CA, USA}
\affiliation[14]{Chan Zuckerberg Biohub Chicago, Chicago, IL, USA}
\affiliation[15]{Department of Chemistry, Northwestern University, Evanston, IL, USA}
\affiliation[16]{Department of Biomedical Engineering, Northwestern University, Evanston, IL, USA}
\affiliation[17]{Robert H. Lurie Comprehensive Cancer Center, Northwestern University, Evanston, IL, USA}
\affiliation[18]{Departments of Pathology and Family Medicine, University of Chicago, Chicago, IL, USA}
\affiliation[19]{Toyota Technical Institute at Chicago, Chicago, IL, USA}
\affiliation[20]{Institute for Population and Precision Health, University of Chicago, Chicago, IL, USA}
\affiliation[21]{Department of Bioengineering, Stanford University, Palo Alto, CA, USA}
\affiliation[22]{Department of Pathology, Stanford University, Palo Alto, CA, USA}
\affiliation[23]{Science for Life Laboratory, School of Engineering Sciences in Chemistry, Biotechnology and Health, KTH Royal Institute of \\ \hspace*{2.25em}Technology, Stockholm, Sweden}
\affiliation[24]{Department of Cell Biology, University of Texas Southwestern Medical Center, Dallas, TX, USA}
\affiliation[25]{Division of Molecular Cell and Developmental Biology, School of Life Sciences, University of Dundee, Dundee, United Kingdom}
\affiliation[26]{Current address: Anthropic, San Francisco, CA, USA}

\date{}

\begin{document}
\maketitle

\begin{abstract}
Biology is at the precipice of a new era—one in which artificial intelligence accelerates and amplifies the ability to study how cells operate, organize, and work as part of systems, revealing why disease happens and how to correct it. Organizations across sectors around the world are prioritizing the application of AI to accelerate basic scientific research, drug discovery, personalized medicine, and synthetic biology. However, despite these clear opportunities, scientific data have proven to be a bottleneck, and progress has been slow and fragmented. Unless the scientific community takes a technology-led, community-focused approach to scaling and harnessing data, we will fail to capture this opportunity to drive new insights and biological discovery.

The data bottleneck presents a unique paradox in scientific research. It is increasingly simple to generate huge volumes of data—thanks to expanding imaging datasets and plummeting sequencing costs~\cite{wetterstrand2025sequencing}—but scientists lack standards and tooling for large biological datasets, preventing the integration of generated datasets into a multimodal foundational dataset that will be key to unlocking truly generalizable models of cellular and tissue function. This contradiction highlights two interrelated problems: there is an abundance of data that is difficult to manage, and a lack of data resources with the necessary quality and utility to realize AI's potential in biology.

Science must forge a new collective approach that enables distributed contributions to be combined into cohesive, powerful datasets that transcend individual dataset purposes. Here, we present a technological and data generation roadmap for scaling scientific impact. We outline the opportunity presented by AI, mechanisms to scale data generation, the need for multi-modal measurements, and a means to pool resources, standardize approaches, and collectively build the foundation that will enable the full potential of AI in biological discovery (``AIxBio'').
\end{abstract}

\section*{Well-structured data sources are needed to train models with rich biological understanding}

Virtual Cell Models (``VCMs'') have quickly become a focal point of AIxBio efforts around the world. These efforts aim to use generative models to help scientists move from discrete understanding of molecules to increasingly complex and holistic models of cellular function~\cite{bunne2024virtual}. At the Chan Zuckerberg Initiative, our vision is to build a family of AI models that learn how cells function at molecular, cellular, and systems levels, enabling scientists to predict and manipulate biological trajectories, accelerating the science for curing, preventing, or managing all diseases. This vision requires unprecedented volumes of multimodal data across cellular scales. Thus, the Chan Zuckerberg Initiative and its Biohubs (``CZI'') will pursue three parallel and complementary scientific challenges: to create novel imaging methods to map intricate biological systems at unprecedented scale and detail; to develop new tools to sense and directly measure inflammation in tissues; and to harness the immune system to enable early detection, prevention, and treatment of disease.

Several VCM efforts, thus far based largely on ``unimodal'' public datasets that measure a single biology analyte–DNA, RNA, or protein–have released initial models that help demonstrate the ambition and promise of VCM models~\cite{nguyen2024evo,adduri2025state,pearce2025transcriptformer,jumper2021alphafold,cui2024scgpt,rosen2024universal,theodoris2023transfer}. However, capturing the opportunity presented by VCM models will require vastly larger and more diverse datasets. To tackle biology's most profound challenges, the scientific community will need to build massive-scale, multi-modal, and interoperable data collections. We argue that no single modality or data type will be sufficient for understanding the wealth of molecules in cells, let alone how those cells cooperatively manifest complex functions across space and time.

We and others have begun collecting multi-scale, multimodal measurements of human biology, making it increasingly essential to develop a systematic approach to data management, consistency, and accessibility. We must move beyond simply generating data; we must now build well-structured data resources to enable the training and development of general-purpose models. With those models in mind, several key features should be incorporated into data generation:

\textbf{Speed} - The pace of data generation must reflect the needs of model training.

\textbf{Cost} - The cost of cutting-edge data must allow iterative cycles.

\textbf{Focus} - Rapid and cost-effective data must be targeted toward core biological problems that reflect concerted modeling efforts.

\textbf{Interoperability} - Data must be transformed into a consistent format to enable training or evaluation.

\textbf{Reproducibility} - Standardized quality control measurements must be created to establish data quality.

\textbf{Infrastructure} - Even given an abundance of data in the appropriate format, data must be stored, accessed, and shared in ways that enable streamlined use.

\section*{Community-wide collaborative data generation is needed to enable this opportunity}

These challenges are deeply interrelated and best approached by developing frameworks by which distributed expertise can be directed toward shared problems. Many efforts already demonstrate the value of such approaches~\cite{lander2001genome,regev2017hca}. In biology, community-generated, highly curated datasets such as the Protein Data Bank~\cite{berman2003pdb}, Sequence Read Archive~\cite{leinonen2011sra}, and CZ CELLxGENE~\cite{czi2025cellxgene} already power AI modeling efforts~\cite{jumper2021alphafold,cui2024scgpt,rosen2024universal,theodoris2023transfer}. CZI sees an opportunity to scale this approach by orchestrating a massive, strategic data effort that can be combined with resources from major biomedical organizations. The necessary scale of this endeavor, as well as the broadly beneficial potential of AI-driven biology, requires that the larger science community come together to advance collective and directed data goals.

\section*{A data strategy for modeling virtual cells}

In order to achieve our vision of AI models that capture cell biology across molecular, cellular, and systems levels, CZI will produce vast amounts of data through its scientific challenges, accelerated by community-wide partnerships with industry, academia, and other philanthropies. Rather than building datasets to answer one specific question, we aim to build a multi-scale, multi-modal reference atlas of cell biology across many species, tissues, and resolutions, using different measurement approaches to predict and simulate biological phenomena from molecular interactions to tissue-level dynamics.

Our data strategy is designed to enable a virtuous cycle of model development, evaluation, and iterative refinement of training data. As we develop and test VCMs, their performance will directly inform which modalities to prioritize and how to balance data collection efforts across different biological scales and systems. Our data generation efforts are structured around three key pillars, each designed to create the basis of the foundational models of the Virtual Cell:

\textbf{Pillar 1: Cellular Diversity and Evolution.} To build a universal model of cell biology, scientists need to understand the vast diversity of cell types across and within organisms. Evolution serves as a vital lens into the different types of viable cellular pathways, functions, states, and types, and CZI is generating a foundational dataset of 100 million cells from 25 diverse animal species to understand how new cell types and their underlying gene regulatory networks evolve. This effort, part of CZI's Billion Cell Project, focuses largely on single-cell transcriptomic data with multiome data for select species, which can be combined with robust existing scRNA-seq datasets on more than 20 organisms, resulting in an overall dataset of 45 organisms. This will enable the critical translation from model organisms to human biology.

\textbf{Pillar 2: Genetic and Chemical Perturbations.} A key function of VCMs will be their ability to model how cells respond to perturbations, in order to ultimately design interventions to reprogram the cell. To build a virtual cell that can predict the effects of genetic and chemical variations, as a function of its initial state and type, CZI will generate three large-scale datasets measuring natural genetic variation, induced genetic variation, and response to chemical perturbations. Our roadmap includes a plan to characterize intrinsic genetic variation in humans and mice by generating data from 100 million cells from 10,000 human donors and 100 mice. This data, consisting of scRNA-seq with matched Whole Genome Sequencing (WGS), will allow us to train models that can predict how a specific perturbation will change a cell's state. To expand our understanding of cell function, we will also generate data from 250 million cells with CRISPR-induced genetic perturbations across a diverse range of cell types and phenotypic states, and measure the effects of small molecule drugs on 70 million cells. This can be combined with existing small-molecule perturbation datasets like Tahoe 100M~\cite{zhang2025tahoe} and smaller perturbation datasets to produce a 400 million cell reference atlas of cellular perturbations.

\textbf{Pillar 3: Multi-scale Imaging and Dynamics.} Cells are highly sophisticated three-dimensional dynamic systems made up of many types of molecules and subcellular components. Cells also interact with one another to form tissues that change, adapt, and respond to their environment. To understand how cells interact across scales from subcellular to multicellular systems, CZI is developing and applying imaging methods that capture dynamic processes across multiple scales. At the molecular level, we are using cryo-electron tomography (cryoET) to visualize individual molecular interactions and assemblies within their native cellular context. This is complemented by systematic sub-cellular measurements of protein localization, building on foundational datasets like OpenCell~\cite{cho2022opencell} and the Human Protein Atlas~\cite{thul2017subcellular}, to provide a dynamic map of the proteome. We also believe optical pooled screens will provide a key resource for modeling the links between molecular perturbation and cell phenotype~\cite{gu2025crispRmap}. At the tissue level, we are leveraging volumetric light-sheet microscopy to image populations of cells within their native tissue context, allowing us to capture real-time observations of tissue architecture and cellular processes. This multi-scale imaging approach provides an essential bridge between molecular readouts and visually observable cellular phenotypes.

Given that the CZI data strategy will generate an unprecedented quantity and diversity of data for AI model innovation and discovery, model benchmarking and interpretability is a key supporting element of these three pillars. The sheer complexity of phenotypic space is a confounding factor for model improvement, so a rich understanding of data-model interactions is critical to develop effective predictive models that will lead to impactful biological discoveries. Our aim is to develop and disseminate an innovative and comprehensive set of benchmarks and evaluation tools to characterize and quantify predictive performance across tasks, cell states, and input data. Because biological datasets represent a composition of many distinct signals, most of which are tangential to any given task, it is critical to map model performance across all components to understand task-specific performance. In addition, we will carefully compare baseline machine learning model performance to understand the exact areas in which transfer learning, enabled by large parameter AI models, can make a transformative impact in performance. These tools will give unprecedented visibility into the behavior of AI models that operate on biological data, enhancing model development and ultimately leading to far more impactful models both within and external to the CZI organization.

CZI views these projects as a starting point. Each has been specifically designed to measure ``anchor variables'' in key biological domains across pillars. For example, most experiments will measure a shared tissue type. This creates biological bridges across datasets that enable model transfer learning. Initial data streams will connect to large-scale data generation efforts targeted towards future biological questions and use of other assay types. We hope this iterative approach will enable cumulative progress towards cohesive VCMs, while generating numerous opportunities for collaborative science and partnership.

\section*{Diverse, integrated data streams will be needed to understand cellular function}

In the field of vision-language models, combining the modalities of natural language and image data has produced dramatic improvements in AI model capability~\cite{radford2021clip}. We hypothesize that the same is true for multimodal scientific data, and that combining large, diverse datasets with AI will unlock new insights into biology.

Our efforts will combine complementary data generation using sequencing, imaging, and mass spectrometry measurements. RNA and DNA sequencing are mature, commercially available and scalable assays that enable broad readouts of genetic potential and cellular states. Dynamic imaging enables the capture of real-time observations of tissue architecture and cellular processes, offering unprecedented opportunities to measure how cells change and respond to biological cues in vivo. Finally, mass spectrometry enables systematic identification and quantification of proteins, post-translational modifications, metabolites, and other molecular species that are invisible to genomic and transcriptomic sequencing technologies.

Organismal potential is fundamentally defined by the genome, while cellular potential is determined by the epigenetic state. The actions of cells, in turn, are executed through proteins. Therefore, models that rely on any single type of measurement—whether transcriptomic, epigenetic, or proteomic—are inherently limited in their ability to model cell function and interaction. Instead, integrating data streams across modalities produces more robust and comprehensive insights. Examples include OpenCell~\cite{cho2022opencell}, which systematically images endogenously-tagged proteins in living cells and, in parallel, identifies their interactions through proteomics to build a dynamic map of the proteome, and Zebrahub~\cite{lange2023zebrahub}, which combines wide-ranging genomic datasets with live imaging of zebrafish development, illustrating the capability to assemble different types of data within a unified framework.

\section*{Data standards will accelerate collective progress}

\begin{figure}[ht]
\centering
\includegraphics[width=\textwidth]{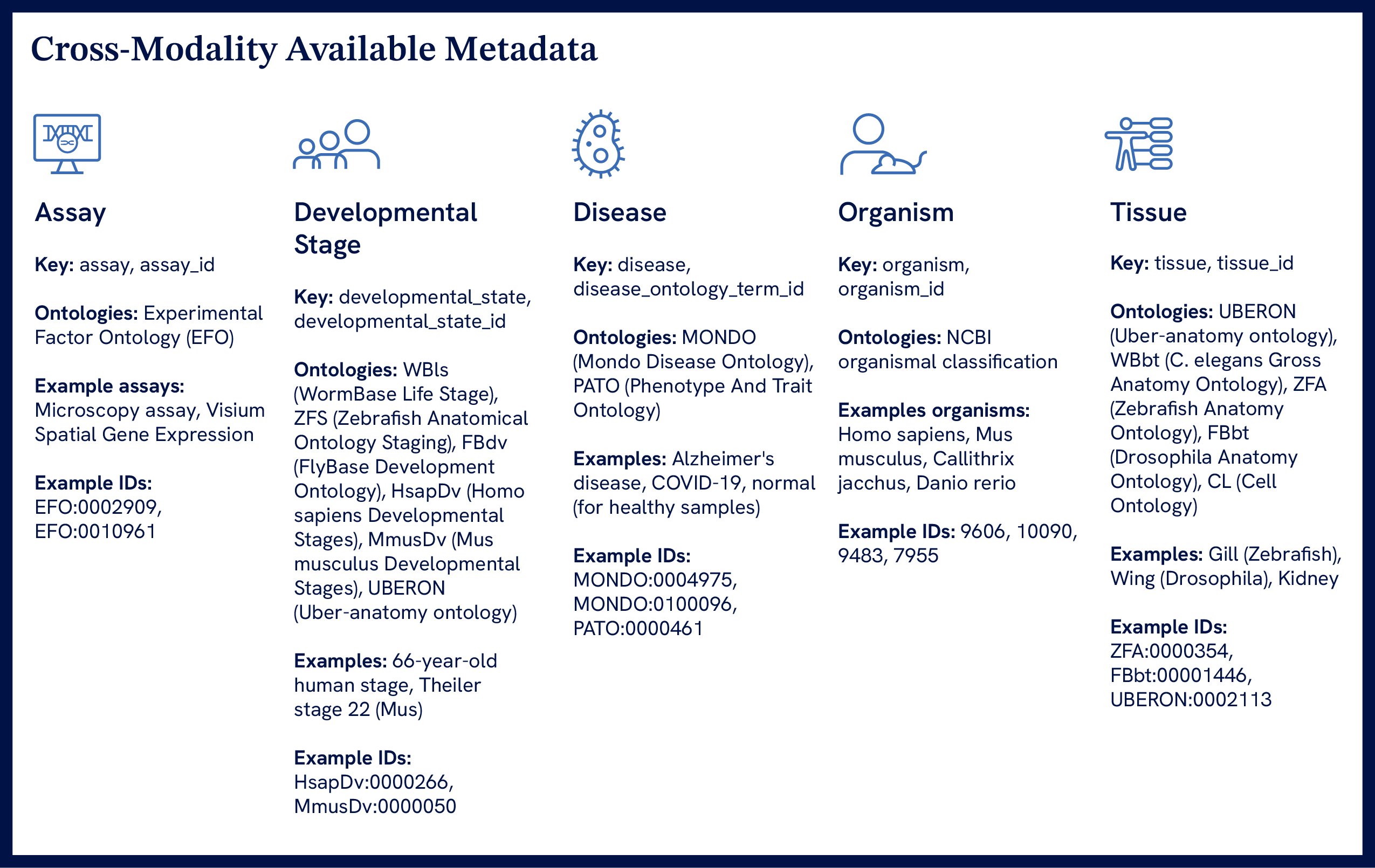}
\caption{Cross-modality Data Standards}
\label{fig:data_standards}
\end{figure}

Integrating and using multi-modal datasets at scale requires consistent, machine-readable metadata and standard formats with efficient toolchains. CZI is creating and supporting standards to bridge biological measurement modalities while preserving flexibility and methodological innovation. Our standards draw from widely used biological standards and ontologies~\cite{regev2017hca,czi2025cellxgene,bard2005ontology,dai2021proteomics} that enable datasets to be easily identified based on shared experimental or biological search parameters, and seamlessly leveraged for cross-modal analyses. To ensure interoperability with the ML community, datasets leverage the croissant standard~\cite{akhtar2024croissant}, which is rapidly gaining popularity and traction. Finally, individual measurements are stored in community standard formats, such as mzML~\cite{martens2011mzml} (mass spectrometry), OME-Zarr~\cite{moore2023omezarr} (imaging), and AnnData~\cite{virshup2021anndata} and TileDB-SOMA~\cite{tiledb2025soma} (sequencing). We use semantic versioning to communicate the impact of changes to our standards and create clarity about their level of maturity for general adoption. The CELLxGENE scRNA-seq data standard~\cite{cellxgene2025standard} is fairly stable and mature, while our other standards (cross-modality~\cite{crossmodality2025standard}, mass spectrometry~\cite{ms2025standard}, imaging~\cite{imaging2025standard}) are evolving more rapidly as the needs of emerging AI modeling approaches are clarified.

By adopting community-developed standard formats, scientists are able to leverage powerful toolchains built to manipulate these datasets at scale. Tiled file formats such as OME-Zarr and TileDB-SOMA accelerate n-dimensional data visualization and efficient slicing and sampling of complex datasets, supporting exploratory data analysis and model training at scale. Drawing inspiration from resources such as the IDR~\cite{williams2017idr}, EMPIAR~\cite{iudin2023empiar}, and Bioimage Archive~\cite{hartley2022bioimage}, CZI is designing a schema to enable federation with community repositories, rather than duplicate their efforts.

Our approach separates experimental metadata from raw formats, establishing the foundation for cross-modal integration across biological measurement modalities. Critically, this approach accelerates data availability to the community: interested researchers can access and explore datasets immediately upon release, rather than waiting for us to standardize the data. While these initial releases may require more effort to work with, they provide early access to valuable data that would otherwise remain locked away during the standardization process. We then iteratively enhance these datasets over subsequent quarters, migrating them to standardized formats that increase interoperability and reusability across modalities, ultimately achieving full FAIR data standards~\cite{fair2025principles}.

All CZI's data standards and pipelines are open-sourced so they can be adapted or extended by the community. This ethos builds on our track record in open science. Tools like CELLxGENE~\cite{czi2025cellxgene} and the CryoET Data Portal~\cite{ermel2024cryoet} have already made it easier for scientists worldwide to explore and visualize biological data, and they underscore our commitment to creating FAIR (Findable, Accessible, Interoperable, and Reusable) data resources that can seamlessly integrate structural, imaging, and molecular data.

\section*{The scale of AI data requires a distributed approach to data management and access}

\begin{figure}[ht]
\centering
\includegraphics[width=\textwidth]{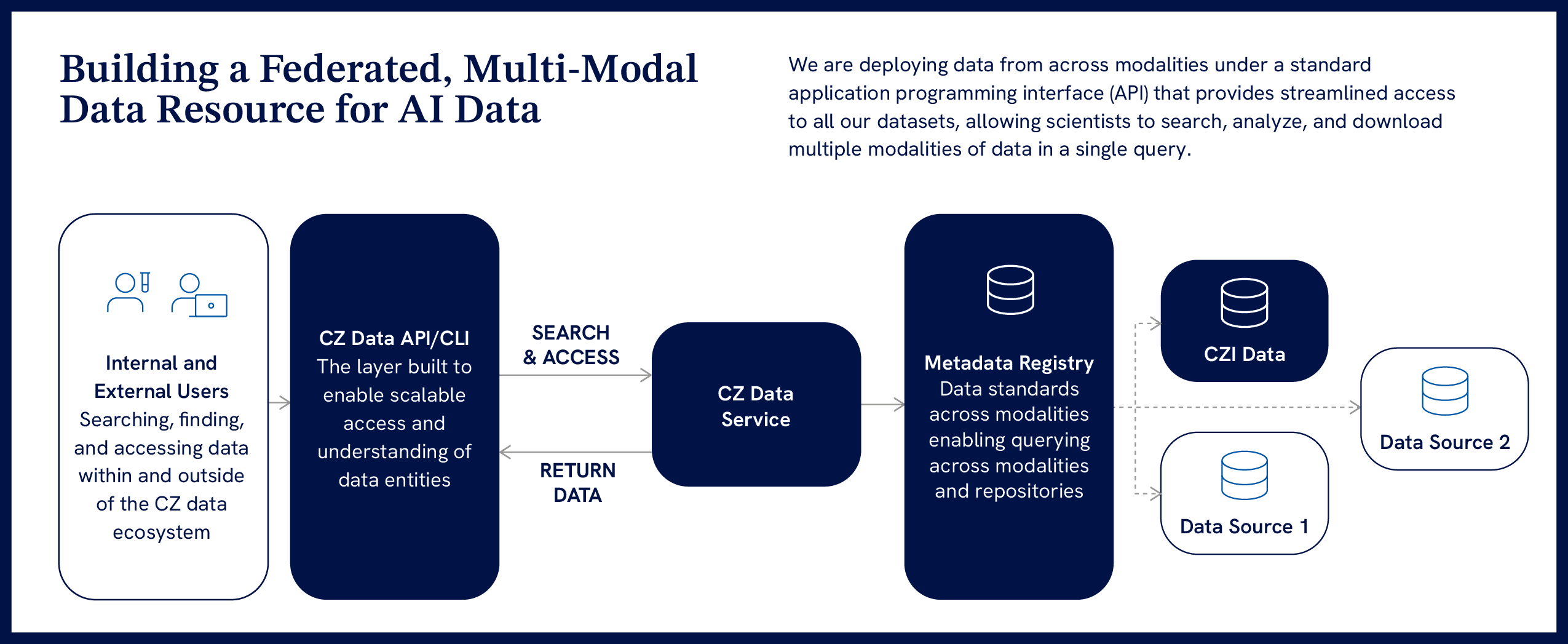}
\caption{CZI Federated Data Service Architecture}
\label{fig:federated_architecture}
\end{figure}

Large-scale imaging efforts, such as light-sheet microscopy, connectomics, and spatial omics, routinely produce gigabyte- to terabyte-scale datasets per experiment, and centralized resources like the BioImage Archive already host petabyte-scale repositories. We estimate that global biological data generation exceeds tens of petabytes annually across multiple major research hubs, with projections indicating cumulative volumes reaching exabyte scale in the coming years. Incorporating publicly available data into AI workflows thus realistically involves managing multi-petabyte data volumes.

Given the scale, heterogeneity, and geographic distribution of data, a federated architecture offers both strategic and logistical advantages. While the use of hierarchical storage where rarely used data is migrated to archival tiers like Glacier Deep Archive provide some cost reduction, centralized aggregation of petabyte-scale imaging datasets would incur unsustainable storage costs. In addition, significant egress charges are incurred when data is downloaded from cloud platforms to external sites. For example, cloud providers like AWS charge $\sim$\$50,000 or more per PB, depending on usage. This cost model disproportionately impacts collaborative and AI-driven workflows that require frequent cross-institutional or compute-layer access to large datasets.

In contrast, a federated model enables data to remain at its source institution while being made accessible through standardized, FAIR-compliant metadata schemas and associated Application Programming Interface (API) or Command Line Interface (CLI)~\cite{bagheri2022quantitative,bajcsy2025enabling}. A strong precedent for this approach has been set by the European Genome-Phenome Archive, which has successfully connected seven national nodes for several years and provided a key resource for the global community~\cite{daltri2025federated}. While data federation comes with a significant coordination burden—as all participating sites must ensure robust performance and high availability—it minimizes data duplication and significantly lowers the cost of large-scale data efforts.

\section*{A federated, multi-model data resource will maximize model development and scientific progress}

Building VCMs alone will be slower, more expensive, and less powerful than collective action. By producing large, rigorously annotated datasets in open formats, CZI aims to not only accelerate our own multi-omic modeling, but to increase adoption of formats and standards across the scientific community, boost the use of biological data, and drive collective progress towards multi-modal modeling.

CZI plans to incorporate existing large publicly available databases—such as those hosted in resources like the Sequence Read Archive (SRA)~\cite{leinonen2011sra}, the European Nucleotide Archive (ENA)~\cite{leinonen2011ena}, and CELLxGENE~\cite{czi2025cellxgene}—to enrich modeling efforts. In return, CZI is committed to giving back to the broader community. All data generated by CZI and its Biohubs will be made as openly accessible as possible (subject to necessary regulatory constraints) and released within a short timeframe to support collective scientific progress.

With this paper, CZI is inviting an alpha cohort to test a CLI aimed at streamlining access to our federated data collection, allowing scientists to search, analyze, and download multiple modalities of data in a single query. Ultimately, this will enable powerful use cases. A scientist, for example, might search for healthy liver datasets across mammalian species, restricting results to studies that include both sequencing and mass spectrometry assays. An ML researcher aiming to train a foundation model can retrieve all measurements of a particular biological observation mode, such as protein expression, across assay types (mass spectrometry, imaging, CITE-seq, etc.). Scientists investigating Alzheimer's disease may query for relevant patient samples alongside in vivo models in various organisms, enabling them to unify transcriptomic, proteomic, and imaging data of disease progression. Even for broad developmental questions—for instance, how prenatal development in invertebrates compares to vertebrate systems—this CLI will offer rapid cross-study integration.

Our current offering is a work in progress. It establishes findability through the cross-modality schema and accessibility through our API and CLI. CZI is committed to expanding the scope of the data model to interface with more data modalities, and to migrating datasets not currently in canonical format standards to establish full interoperability, therefore maximizing reusability. This approach enables us to balance data quality and release velocity. We will use this approach to make billions of cellular measurements publicly available in the coming years.

We believe the key to improving the value of a public resource like this is to engage with the scientific community at every step. We welcome feedback, partnerships, and data contributions, confident that a truly collaborative, open-access data environment will further enrich the value for users of the CLI and yield richer query results. Researchers interested in participating in the alpha cohort testing our CLI can sign up at this \href{https://chanzuck.co1.qualtrics.com/jfe/form/SV_dmdfn4ZyxYHBvMy?Q_CHL=preprint2025Oct}{Data Tools Interest/Feedback Form}.

Ultimately, this technological blueprint seeks to empower the broader scientific community. By combining multiple forms of data on an unprecedented scale and creating accessible tools to analyze that data, we hope to advance fundamental biology and accelerate the discovery of novel therapies. Whether it's through a deeper understanding of how cells function, breakthroughs in sensing inflammation, or the design of targeted immunotherapies, our ultimate goal is a healthier future enabled by open data and collaborative science.

\bibliographystyle{unsrtnat}
\bibliography{references}

\end{document}